\documentclass[iop,apj]{emulateapj}

\usepackage{subfigure}

\bibpunct[; ]{(}{)}{,}{a}{}{;}

\def\simgt{\lower.5ex\hbox{$\; \buildrel > \over \sim \;$}}
\def\simlt{\lower.5ex\hbox{$\; \buildrel < \over \sim \;$}}

\def\etal{{et~al.}}
\def\amin{\ifmmode^{\prime}\else$^{\prime}$\fi}
\def\asec{\ifmmode^{\prime\prime}\else$^{\prime\prime}$\fi}

\def\simgt{\lower.5ex\hbox{$\; \buildrel > \over \sim \;$}}
\def\simlt{\lower.5ex\hbox{$\; \buildrel < \over \sim \;$}}

\newcommand\xte{{\it RXTE\/}}

\newcommand\chandra{{\it Chandra}}

\newcommand\hess{{H.E.S.S.}}

\newcommand\snr{G76.9$+$1.0}
\newcommand\cxo{CXOU~J202221.68$+$384214.8}
\newcommand\cxos{CXOU~J202221}
\newcommand\psr{PSR~J2022$+$3842}
\newcommand\psrb{PSR~J1617$-$5055}
\newcommand\pasa{PASA}

\slugcomment{Accepted for publication in The Astrophysical Journal}

\shorttitle{An Energetic Pulsar in SNR \snr}
\shortauthors{Arzoumanian \etal}

\begin{document}

\title{Discovery of an Energetic Pulsar Associated with SNR \snr}

\author{Z. Arzoumanian\altaffilmark{1,2}}
\email{Zaven.Arzoumanian@nasa.gov}

\author{E. V. Gotthelf\altaffilmark{3}}

\author{S. M. Ransom\altaffilmark{4}}

\author{S. Safi-Harb\altaffilmark{5}}

\author{R. Kothes\altaffilmark{6}}

%\and{}

\author{T. L. Landecker\altaffilmark{6}}

\altaffiltext{1}{Center for Research and Exploration in Space
Science and Technology and X-ray Astrophysics Laboratory, NASA
Goddard Space Flight Center, Code 662, Greenbelt, MD 20771, USA}

\altaffiltext{2}{Universities Space Research Association, Columbia, MD 21044, USA}

\altaffiltext{3}{Columbia Astrophysics Laboratory, Columbia University,
550 West 120th Street, New York, NY 10027, USA}

\altaffiltext{4}{National Radio Astronomy Observatory,
520 Edgemont Road, Charlottesville, VA 22901, USA}

\altaffiltext{5}{Canada Research Chair. Department of Physics and Astronomy, University of
Manitoba, Winnipeg, MB, R3T 2N2, Canada}

\altaffiltext{6}{National Research Council of Canada, Herzberg
Institute of Astrophysics, Dominion Radio Astrophysical Observatory,
Box 248, Penticton, BC, V2A 6J9, Canada}

%\altaffiltext{7}{Department of Physics and Astronomy, University of
%Calgary, Calgary, AB, Canada}

%\slugcomment{DRAFT, \today}

\begin{abstract}

We report the discovery of \psr, a 24 ms radio and X-ray pulsar in the
supernova remnant \snr, in observations with the
\chandra\ X-ray telescope, the Robert C. Byrd Green Bank Radio
Telescope, and the {\it Rossi X-ray Timing Explorer\/} (\xte). The pulsar's 
spin-down rate implies a rotation-powered luminosity $\dot E =
1.2\times 10^{38}$~erg~s$^{-1}$, a surface dipole magnetic field 
strength $B_s = 1.0\times 10^{12}$~G, and a characteristic age of 8.9~kyr.
\psr\ is thus the second-most energetic Galactic pulsar known, after
the Crab, as well as the most rapidly-rotating young, radio-bright
pulsar known. The radio pulsations are highly dispersed and
broadened by interstellar scattering, and we find that a large
($\delta f / f \approx 1.9 \times 10^{-6}$) spin glitch must have
occurred between our discovery and confirmation observations. The
X-ray pulses are narrow (0.06 cycles FWHM) and visible up to 20 keV,
consistent with magnetospheric emission from a rotation-powered
pulsar. The \chandra\ X-ray image identifies the pulsar with a hard, unresolved source
at the midpoint of the double-lobed radio morphology of \snr\ and
embedded within faint, compact X-ray nebulosity. The spatial
relationship of the X-ray and radio emissions is remarkably similar
to extended structure seen around the Vela pulsar. The combined
\chandra\ and \xte\ pulsar spectrum is well-fitted by an absorbed
power-law model with column density $N_{\rm H} = (1.7\pm0.3)
\times10^{22}$~cm$^{-2}$ and photon index $\Gamma = 1.0\pm0.2$; it
implies that the \chandra\ point-source flux is virtually 100\%
pulsed.  For a distance of 10~kpc, the X-ray luminosity of \psr\ is
$L_{\rm X}(\mbox{2--10~keV}) = 7.0 \times 10^{33}$~erg~s$^{-1}$.
Despite being extraordinarily energetic, \psr\ lacks a bright X-ray
wind nebula and has an unusually low conversion efficiency of
spin-down power to X-ray luminosity, $L_{\rm X}/\dot E = 5.9 \times
10^{-5}$.

\end{abstract}

\section{Introduction}
\setcounter{footnote}{0}

The past decade has been witness to a highly productive
back-and-forth relationship between X-ray and radio studies of
supernova remnants (SNRs). Their nonthermal radio emissions,
serving as tracers of particle acceleration, have motivated
follow-up X-ray observations, most notably with the \chandra\
telescope. In many instances, X-ray imaging has revealed point-like
sources surrounded by extended synchrotron structures, typically
axisymmetric; these pulsar wind nebulae (PWNe) arise from the
shocked outflows of relativistic particle winds driven by a central
engine, the powerful time-varying fields in a neutron star
magnetosphere. Targeted radio periodicity searches of PWNe
have then uncovered previously unseen pulsations, providing the spin
periods and, eventually, estimated ages, magnetic field strengths,
and spin-down luminosities of the neutron stars, without which
a meaningful physical understanding of the surrounding phenomena
would be beyond reach. A few specific examples of this general
picture of radio/X-ray symbiosis have been the discoveries of the
pulsars in the remnants 3C58, G54.1+0.3, G292.0+1.8, and in the
``Mouse'' \citep[see, e.g.,][]{2004IAUS..218...97C}.

The Crab Nebula has been viewed for decades as a prototype, but
there is growing evidence of PWNe with substantially different
properties. At least three objects, among them the Vela SNR, have nonthermal
radio spectra that are steep in comparison to the Crab (i.e., with
spectral indices $\alpha\approx0.6$, where flux
$S_{\nu}\propto\nu^{-\alpha}$ at frequency $\nu$) and morphologies characterized
by radio emission lobes straddling a central
depression. While the nature of one such candidate PWN, DA~495, %as an SNR
was once in question, X-ray imaging observations
led to the discovery of an X-ray nebula and a presumed neutron star,
both residing near the radio ``hole''
%\citep{2008ApJ...687..505A,2008ApJ...687..516K}. 
\citep[hereafter, ASL+08]{2008ApJ...687..505A}.
Radio and X-ray
pulsation searches have not, however, detected a pulsar signature,
so a key ingredient is lacking from our understanding of DA~495. In
this paper, we describe a similar series of observations directed
toward understanding the third member of this group of unusual SNRs,
\snr. 

\begin{figure*}[t]
\centerline{
\subfigure{\includegraphics[width=\columnwidth,angle=270]{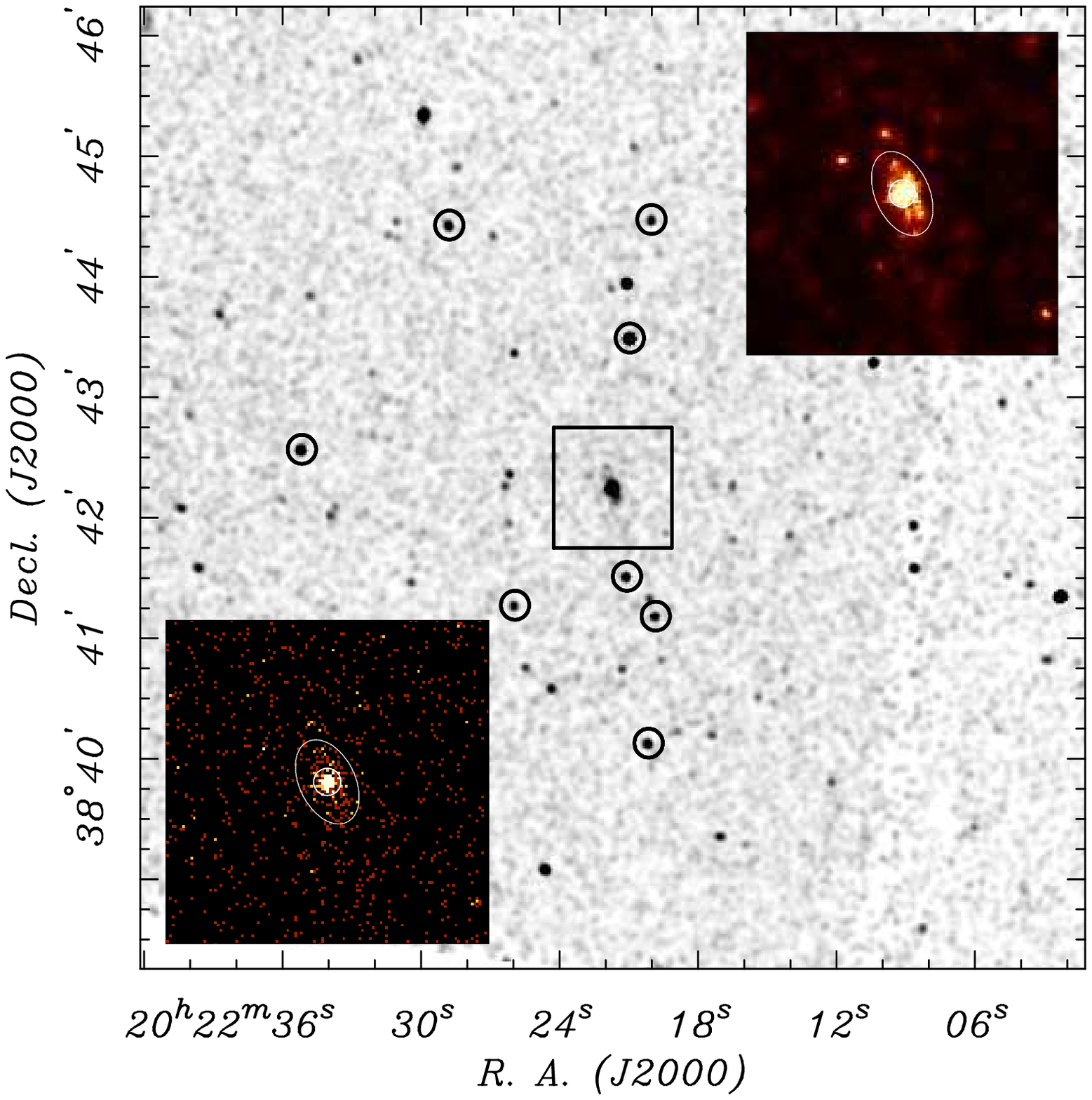}}%
\hfill
\subfigure{\includegraphics[angle=270,width=\columnwidth]{g769.vla21_color.ps}}%
}
\caption{{\it Left:\/} \chandra\ ACIS-S3 0.5--7~keV X-ray image of
the center of \snr, smoothed, exposure-corrected, and scaled to emphasize
the field point sources. The $1\arcmin\times1\arcmin$ box is centered on \cxo, while
the circled sources were used to register the image. {\it Lower
inset:\/} The unsmoothed count-rate map within the box
containing the candidate neutron star and PWN. {\it Upper inset:\/}
The result of adaptive smoothing of the same image. Overplotted are
the regions used to extract spectral information for the unresolved
source ({\em circle}) and diffuse emission ({\em ellipse}). {\it Right:\/} 
$1.49$~GHz map of SNR \snr\ obtained using
the VLA (see \citealp{lhw93} for details) with the X-ray pulsar and
nebula regions superposed ({\em contours}). The pulsar
lies along the ridge of emission connecting the two lobes, adjacent
to the central ``hole.'' The arrangement is strikingly similar to
the radio/X-ray configuration of the Vela pulsar and PWN (see
Figure~10 of \citealp{dlm+03}) as well as of DA~495 (Figure~1 of 
%\citealp{2008ApJ...687..505A}
ASL+08).
}
\label{fig:chandraimage}
\end{figure*}

SNR \snr\  was discovered in a 408~MHz DRAO survey of the Cygnus~X 
region and initially characterized as a ``possible galactic object''
\citep{whl91}. Follow-up three-band radio observations using the Very
Large Array (VLA) by \cite{lhw93} resolved a two-lobed structure $4'$
in size, with a connecting bridge of emission, embedded in a faint,
roughly circular emitting region $9'\times12'$ across. Based on its steep,
nonthermal spectrum, polarization, and morphological similarities to
center-filled SNRs (in important respects such as the lack of an outer 
boundary and radially decreasing flux away from the central depression), 
these authors favored an SNR interpretation. In this picture, the radio
 lobes trace the Crab-like synchrotron emission of an evolved pulsar 
 wind nebula. 
 
Here we present imaging, spectral, and timing studies of \snr\ in
X-rays and radio, culminating in the discovery of \psr\  
%(the pulsations of which are visible in both bands) 
and of its associated X-ray PWN. In \S\ref{sec:chandra}, we describe
analysis of a \chandra\ observation that yielded the first
identification of the SNR's central engine, a neutron star with its
surrounding nonthermal nebula.  In \S\ref{sec:radio}, we describe the
discovery of the radio pulsar J2022+3842 and follow-up radio timing
observations with the Green Bank Telescope (GBT) that have provided 
its spin-down rate and pulse 
properties. In \S\ref{sec:xte}, we present the results of a deep
{\it Rossi X-ray Timing Explorer} (\xte) observation: 
we show that the pulsed X-ray flux is
consistent with the \chandra\ point-source flux, cementing the pulsar's
association with the SNR.  

For reasons developed below, we adopt a distance to \snr\ of 10~kpc
throughout this work. %\citep{lhw93}.

\section{\chandra\ Observations and Results}
\label{sec:chandra}

SNR \snr\ was observed for 54~ks by the \chandra\ observatory
on 2005 August 01 UT using the Advanced CCD Imaging Spectrometer
\citep[ACIS;][]{bur97} operating in the full-frame TIMED/VFAINT
exposure mode (ObsID \#5586). This detector is sensitive to X-rays in
the 0.3--12~keV energy range with a resolution of $\Delta E /E \sim
0.06$ FWHM at 1~keV. The CCD pixel scale is $0.5\arcsec$,
comparable to the telescope's on-axis spatial resolution. Low-level
data were reprocessed with the latest
calibrations using the CIAO script {\tt chandra\_repro}. The data
analysis made use of the CIAO (V4.3), FTOOLS (V6.10), CALDB (V4.4.2),
and XSPEC (V12.6.0q) software packages. A total of 53.7~ks of
live-time was accumulated.
No additional time filtering was necessary as the background rate was
stable over the course of the observation.  We followed the CIAO
online science threads to create ACIS spectra and images. 
To allow for imaging with the best available angular
resolution, we used the {\tt EDSER} reprocessing
option, applying an energy-dependent sub-pixel event-repositioning 
algorithm to each photon \citep{2004ApJ...610.1204L}.

The left-hand panel of Figure~\ref{fig:chandraimage} presents the
exposure-corrected image, in the 0.5--7~keV band, acquired by the
ACIS-S3 and S2 CCDs, which together covered the entire extent of
the radio SNR. No morphological
evidence of \snr\ itself is apparent nor is there any hint of
resolved diffuse emission in images created in the soft ($<2$~keV)
or hard ($>2$~keV) energy bands; instead, several faint point-like
sources are detected.  The brightest of these, enclosed within a box
in the figure, is located close to the optical axis and contains
$N=1190$ photons. This source is not associated with any known
stellar counterpart or object in the 2MASS image of the field. Its
identification as a possible neutron star associated with \snr\ is
reinforced by the presence of faint but distinct X-ray nebulosity
surrounding a brighter, unresolved core. To highlight the diffuse
emission, we show as insets in Figure~\ref{fig:chandraimage} both
the raw and smoothed 0.5--7 keV count-rate maps for the boxed
$1\arcmin\times1\arcmin$ region centered on the source.
The smoothed image was produced using a circular convolution kernel
of varying size, requiring a minimum of 5 counts within the kernel
diameter.
Although the putative nebula is faint,
there is a clear excess of emission elongated on a line with
position angle $\simeq 26^{\circ}$ North through East. This and the 
relative faintness of the point source argue against identification of the 
diffuse emission as a dust-scattering halo; these are normally circular
and found around much brighter sources.

The $1.49$~GHz Very Large Array (VLA) radio image shown in the
right-hand panel of Figure~\ref{fig:chandraimage} places the X-ray
nebula in the context of the double-lobed radio structure. 
The unambiguous alignment of the long axis of the diffuse X-ray
emission with the radio ridge connecting the lobes 
suggests that the radiation processes underlying 
the emissions in both bands are related, such as in 
the canonical picture of synchrotron emission
from relativistic particles originating in the pulsar's magnetosphere.

To refine the location of the putative pulsar, the X-ray image was registered 
using the coordinates of eight 2MASS infrared counterparts obtained
from the NASA/IPAC Infrared Science Archive. The updated centroid is
located at (J2000.0) R.A. = $20^{\rm h}22^{\rm m}21\fs689$,
Decl.\ = $+38^{\circ}42^{\prime}14\farcs82$ with a $1\sigma$
uncertainty of ($0\farcs09,0\farcs07$) in the two coordinates,
respectively.  The required shift in R.A. and Decl.\ was
($-0\farcs012,-0\farcs162$).  This source, \cxo, is hereafter referred to
by the shortened name \cxos.

We extracted a spectrum for \cxos, and generated an appropriate response
matrix, with the {\tt specextract} CIAO script, using a circular $2\farcs5$-radius 
aperture (representing $>90$\% encircled energy, depending on photon 
energy) centered on the source.  The diffuse emission
and cosmic and detector backgrounds contribute negligibly ($\sim 10$
counts) in this region.  With a maximum count rate in a pixel of $<
4\times10^{-3}$~s$^{-1}$, photon pile-up could safely be ignored. The 
0.5--10 keV
spectrum was grouped with a minimum of 15 counts per spectral channel
and fitted using XSPEC to an absorbed power-law model, a
common property of young, rotation-powered pulsars. (Blackbody models
are formally allowed, but produce unnaturally high temperatures.) An excellent
fit, with $\chi^2=69.5$ for 70 degrees of freedom (DoF) was obtained
for an absorbing column $N_{\rm H} = (1.6\pm0.3) \times
10^{22}$~cm$^{-2}$ and photon index $\Gamma = 1.0\pm0.2$ (90\%
confidence intervals are used throughout). The absorbed 2--10~keV
source flux is $F_{\rm PSR} = 5.3 \times
10^{-13}$~erg~cm$^{-2}$~s$^{-1}$, implying an isotropic
luminosity of $L_{\rm X} = 7.0 \times 10^{33}$~erg~s$^{-1}$ at
10~kpc distance. As described in \S\ref{sec:xtespec}, this \chandra\ spectrum
was also used in joint fits with \xte\ data.

%To investigate the spectrum of the nebula, we used 
%an elliptical extraction region (Figure~\ref{fig:chandraimage}), with
%semi-major and -minor axes of $8\farcs3 \times 5\farcs1$,
%centered on the pulsar, but with the point-source region excluded. An
%annular region also centered on the pulsar but well outside the nebular 
%extent, with inner and outer radii of $20\arcsec$ and $40\arcsec$ respectively, 
%was used to estimate the local background. A total of $N=84\pm13$ 
%background-subtracted counts could be ascribed to the nebula.
%Its faintness did not allow a well-constrained spectral
%fit, but the nebula is clearly a hard source and there are indications that 
%a fit to a power-law model would yield a fairly flat index 
%$\Gamma\sim1$. Nevertheless, to more firmly estimate the nebular flux, we fixed the 
%column density to the point-source best-fit value and assumed a nominal 
%absorbed power-law model with $\Gamma = 1.4$, a value derived from the 
%spectral index trend between energetic pulsars and their nebulae 
%reported by Gotthelf (2004). 
%The absorbed 2--10~keV flux for the
%putative PWN is $F_{\rm PWN} \approx 4 \times
%10^{-14}$~erg~cm$^{-2}$~s$^{-1}$. The flux ratio $F_{\rm PWN}/F_{\rm
%PSR} \approx 0.08$ is well below that expected for similarly
%energetic pulsars (Gotthelf 2004). The uncertainty in this result is
%dominated by the small number of nebula photons and not the choice
%of $\Gamma$.

To investigate the spectrum of the nebula, we used an elliptical
extraction region (Figure~\ref{fig:chandraimage}), with semi-major
and -minor axes of length $8\farcs3 \times 5\farcs1$, centered on
the pulsar, with the point-source region ($r<2\farcs5$) excluded.
For the local background, we extracted photons from a concentric
annular region well outside the nebular extent ($20\arcsec < r <
40\arcsec$).  We also performed a {\tt MARX} simulation to determine the
small (2.6\%) flux contribution within the nebular region from
point-source photons scattered into the point-spread function wings; this
simulation took into account the measured pulsar spectrum. The tally
of background-subtracted counts from the nebula is then $N=76\pm13$.
The faintness of the nebula does not allow a well-constrained
spectral fit; instead we fixed the column density to the best-fit
value for the unresolved source and assumed a nominal absorbed
power-law model with $\Gamma = 1.4$, a value derived from the
empirical trend relating the spectral indices of energetic pulsars
and their nebulae \citep{got04}.  The absorbed 2--10~keV flux for the
putative PWN is $F_{\rm PWN} \approx 4 \times
10^{-14}$~erg~cm$^{-2}$~s$^{-1}$. The flux ratio $F_{\rm PWN}/F_{\rm
PSR} \approx 0.08$ is well below that expected for similarly
energetic pulsars \citep{got04}. The uncertainty in this result is
dominated by the small number of nebula photons and not the choice
of $\Gamma$.

\section{Radio Observations and Results}
\label{sec:radio}
Motivated by the \chandra\ discovery of the point source and possible
PWN at the heart of SNR \snr, we searched for radio pulsations with
the GBT. The Spectrometer SPIGOT data-acquisition 
backend \citep{2005PASP..117..643K} was used for 5
hours on each of two days, with 600 MHz of useful bandwidth centered at 1.95
GHz. The instrument provided 768 channels with 16-bit sampling of
summed polarizations at 81.92~$\mu$s intervals. The PRESTO software
\citep{2002AJ....124.1788R} was used to search for pulsations, and a
candidate 24-ms periodicity was detected on both days with formal
significances of 7$\sigma$ on MJD 54397 and 12$\sigma$ on MJD 54400.
Confirmation observations were carried out on 2009 May 6 (MJD 54957)
using the GUPPI data-acquisition system configured similarly to the
discovery instrumentation, but with the addition of full-Stokes
sensitivity.
The difference in the pulse periods derived at these epochs gave the
first indication of the pulsar's spin-down rate and thence its
canonical age, magnetic field strength, and spin-down power (see
below).
A search for single ``giant'' pulses yielded no compelling
candidates, despite the fact that the pulsar's estimated magnetic
field strength at the light cylinder, suggested to be relevant in
the production of giant pulses \citep{1996ApJ...457L..81C}, is
approximately 70\% that of both the Crab pulsar and PSR~B1937+21,
and comparable to that of PSR~B1821$-$24, all of which exhibit giant
pulses.

\begin{figure}[t]
\hspace{0.1in}
\begin{center}
\includegraphics[height=0.9\linewidth,angle=270,clip=true]{radio_profile_pol_v2.ps}
\includegraphics[width=0.9\linewidth,angle=0,clip=true]{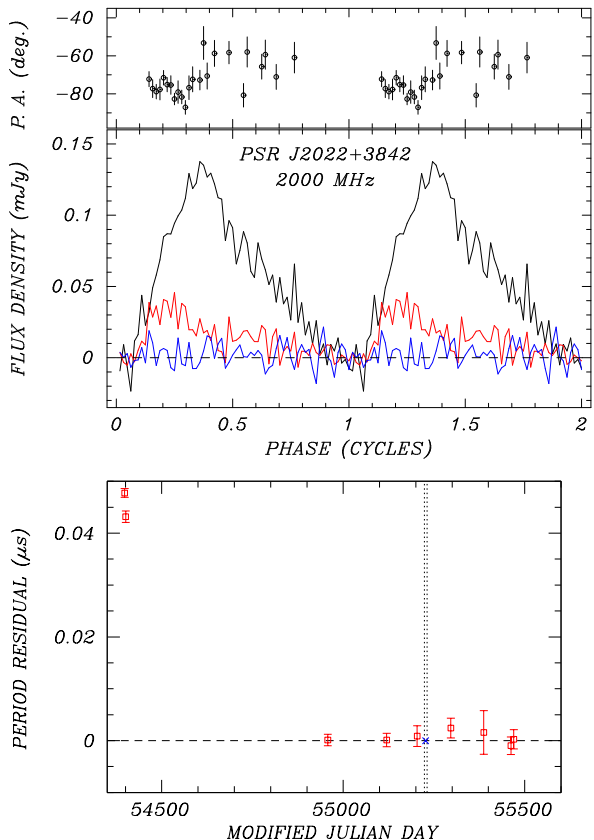}
\end{center}
\caption{{\it Top:\/} Radio pulse properties of PSR J2022+3842
averaged over a 700~MHz band centered at 1.95~GHz, and representing
a total integration time of approximately 24 hours. The evolution
with pulse phase of total ({\em black line}), linearly polarized
({\em red}), and circularly polarized ({\em blue}) flux are shown. 
Phase zero is arbitrary. {\it Bottom:\/} Radio pulse
period evolution ({\em red squares}) with 1$\sigma$ uncertainty
error bars; the blue cross corresponds to the \xte\
observation (uncertainty much smaller than the plotted symbol).
Residuals are relative to the timing model 
shown in Table~\ref{tab:ephem}. 
The pulse period at the discovery epoch of the radio
pulsations, around MJD~54400, was significantly longer than 
at later times, indicating that a spin-up glitch occurred at
some point before the confirmation observation near MJD~54950.}
\label{fig:radiotiming}
\end{figure}

Additional radio observations, also with the GUPPI system, were
carried out at roughly 3-month intervals to constrain the pulsar's
long-term timing behavior. 
%(Figure~\ref{fig:radiotiming}).  
Notably, the pulse periods measured during and after the confirmation
observation differed substantially from those detected during the two
discovery observations made 1.5 years earlier, implying that a spin
``glitch'' of magnitude $\Delta P/P \simeq 1.9\times 10^{-6}$ had
occurred at an unknown epoch in this interval. The spacing of the
radio timing observations together with apparent ``timing noise''
instability in the pulsar's rotation on similar timescales precluded
derivation of a phase-connected timing solution. We therefore
determined the long-term average spin-down rate,
$4.3192(27)\times10^{-14}$, through a least-squares fit to the
multi-epoch period measurements (Figure~\ref{fig:radiotiming}), a 
result consistent with the
short-term X-ray-derived ephemeris shown in Table~\ref{tab:ephem}
(see \S4.1).

The pulsar's flux density at 2 GHz, $S_{\rm 2\,GHz}$, is 60~$\mu$Jy;
its 4.8~GHz flux density of $\approx 45$~$\mu$Jy suggests that \psr\
has an unusually flat spectrum, $\alpha \simeq -0.33$. The implied
radio pseudo-luminosity $S_{\rm 1.4\,GHz}d^2 \simeq 7\,D^2_{10}$~mJy~kpc$^2$ is
unremarkable, low in comparison to the majority of known pulsars,
but an order of magnitude higher than the luminosities of other
young, faint pulsars discovered in deep searches of SNRs
\citep{2004IAUS..218...97C}. Even with a long cumulative exposure,
our best average pulse profile (Figure~\ref{fig:radiotiming})
retains significant statistical noise; despite this faintness, a number of 
relevant pulse shape and polarization results are available. 

The radio pulsations of \psr\ are significantly affected by dispersion and scattering 
due to propagation in the interstellar medium, which are typically important at 
frequencies $\lesssim 1$~GHz, even in our 2~GHz observations. To estimate
dispersion measure (DM) and the pulse broadening timescale $\tau^{\rm scatt}$,
we formed a pulse-shape model assuming a gaussian-shaped intrinsic profile, 
a one-sided exponential-decay impulse response function for the scatter-broadening, 
and a $\nu^{-4}$ frequency dependence 
for the width of the exponential.  We fit the S-band (i.e.,
2\,GHz) data to this model in four evenly-spaced frequency sub-bands, allowing the 
width of the intrinsic gaussian profile to vary and also accounting for 
a DM bias caused by the frequency-dependent pulse broadening 
due to scattering.  This simple model fit the data well, but we caution that the 
assumptions inherent in the model may introduce significant systematic errors, 
perhaps as large as twice the statistical uncertainties quoted below. 
%and indicates 
%that the intrinsic pulse profile is wide ($\sim$30\%)
%, in agreement with 
%our one low signal-to-noise detection of the pulsar at 4.8\,GHz with the 
%GBT.  
%Given the above assumptions, the systematic error on $\tau^{\rm scatt}$ may
%be as large as a factor of two, but the fact that 
%the pulsar is heavily contaminated by scattering at frequencies below 
%$\sim$2.5\,GHz is not in doubt.

The NE2001 Galactic electron distribution model
\citep{2002astro.ph..7156C} fails to accommodate the derived 
${\rm DM} = 429.1\pm0.5$~pc~cm$^{-3}$ in the direction of
\psr---formally, the model suggests a lower bound on the distance to
the pulsar of 50 kpc. Instead, a likely over-density of free
electrons in the Cygnus region, along the line of sight, accounts
for the higher-than-expected dispersion---a similar explanation has
been advanced for the unexpectedly large DM of the
nearby PSR J2021+3651 \citep{2002ApJ...577L..19R}. 

In this direction,
the Perseus Arm lies at a distance of approximately 6~kpc, and the
Outer Arm at $\gtrsim10$ kpc. For convenience in scaling
distance-dependent parameters and because the larger distance brings
X-ray efficiencies more approximately in line with those of other
young pulsars (see~\S\ref{sec:discuss}), we adopt a distance $d = 10\, D_{10}$~kpc. Independent
supporting evidence for the large distance derives from H\,{\sc i}
absorption and X-ray studies of the PWN CTB 87
\citep{2003ApJ...588..852K}, roughly $2\degr$ from \snr\ on the sky.
At a distance of 6.1 kpc, CTB 87 lies in the Perseus arm and its
X-ray-derived absorbing column is $N_{\rm H} =
(1.4\pm0.2)\times10^{22}$~cm$^{-2}$ at 90\% confidence (Safi-Harb, Matheson \&
Kothes, in preparation); although their uncertainties are large, the nominal
H\,{\sc i} column for \snr\ is somewhat larger than that for CTB 87.
A radio bright extragalactic point source adjacent to CTB 87
exhibits a deep absorption component at a velocity of about
$-85$~km~s$^{-1}$ not seen in the absorption profile of CTB 87. This
feature could account for the larger foreground absorption toward
\snr\  if the latter is located at a significantly larger distance,
such as the Outer Arm or beyond it.

The NE2001 model also predicts a timescale for pulse broadening due
to interstellar scattering of 0.3 ms at 1 GHz; instead, we find
$\tau^{\rm scatt}_{\rm 1 GHz} = 55 \pm 7$~ms, which scales to
$3.8$~ms at the center of our observing band but varies
strongly across it. 
The long decay of the pulse, here averaged
over the 700 MHz band, can be seen in the upper half of
Figure~\ref{fig:radiotiming}. Consistent with this result, an
exploratory observation centered at 800 MHz sky frequency failed to
detect the pulsar; similar attempts at 1.4 GHz and 5 GHz yielded
weak detections. At 5 GHz, the pulse was unexpectedly broad, consistent 
with the results of the frequency-subband fit to the pulse shape. Thus,
in addition to the importance of scatter-broadening at the lower
frequencies, the pulse appears to be intrinsically broad, roughly
30\% of the pulse period, full width at half maximum. The
well-established trend of pulse widths increasing as $P^{-0.5}$
underestimates this result by a factor of 2--3 (e.g., for the
outermost conal component of \citealp{1999A&A...346..906M}).

Polarization calibration and analysis using the PSRCHIVE software 
package \citep{2004PASA...21..302H} 
reveals that the radio pulse is substantially linearly polarized, despite
the depolarizing effects of multipath propagation in the scattering
tail, incomplete removal of Faraday rotation (due to the low
signal-to-noise ratio), and other effects. The true polarized
fraction in the pulse is likely to be significantly larger than we have
observed.
The variation of the polarization vector's position angle with
rotational phase, shown in the top-most panel of Figure~\ref{fig:radiotiming}, 
is found to be fairly flat, inconsistent with the canonical rotating-vector
model \citep{1969ApL.....3..225R}, but similar to behavior seen in
the Crab pulsar (especially its high-frequency components; 
\citealp{1999ApJ...522.1046M}) and several millisecond-period ``recycled'' pulsars 
\citep{2004MNRAS.352..804O}. Perhaps not coincidentally, many of these same pulsars 
exhibit nonthermal magnetospheric pulsations in X-rays. Finally, we note
that the pulsar's rotation measure is ${\rm RM} \simeq +270$~rad~m$^{-2}$.

\section{\xte\ Observations and Results}
\label{sec:xte}

The field containing \cxos\ was observed by \xte\ for a
total of 99~ks over the 8-day span 2010 January 27--February
4 UT (observation P95316). The data were collected with
the Proportional Counter Array \citep[PCA;][]{jah96} in the GoodXenon
mode with an average of 1.8 of the 5 proportional counter units
(PCUs) active. In this mode, photons are time-tagged to $0.9$ $\mu$s
and have an absolute uncertainty better than 100 $\mu$s
\citep{rot98}. The effective area of five combined detectors is approximately
$6500$ cm$^{2}$ at 10~keV with a roughly circular field-of-view of
$\sim 1^{\circ}$ FWHM. Spectral information is available in the
2--60~keV energy band with a resolution of $\sim 16\%$ at 6~keV.

Production PCA data for this observation were obtained from NASA's
HEASARC archive and time-filtered using the standard criteria. This
rejected intervals of South Atlantic Anomaly passages, Earth
occultations, and other periods of high particle activity, resulting
in a total of 94.2~ks of useful data. The photon arrival times were
transformed to the solar system barycenter in Barycentric Dynamical
Time (TDB) using the JPL DE200 ephemeris and the \chandra-derived
coordinates of \cxos.

\subsection{Timing Analysis}
\label{sec:xtetim}
The timing analysis was restricted to PCA channels 2--50,
corresponding approximately to the 2--20~keV energy range, and from the
top Xenon layer of each PCU, to optimize the signal-to-noise for a
typical pulsar spectrum. Because of the long observation span,
photon detection times were accumulated in 4-ms bins and a
$2^{28}$-point accelerated Fast Fourier Transform (FFT) was
performed to search for a periodic signal, testing a range of
plausible frequency derivatives.
This immediately revealed a signal at 24~ms of sufficient strength
to analyze short intervals of data and build up a fully
phase-coherent timing solution using the time of arrival (TOA)
method, as follows.  

\begin{figure}[t]
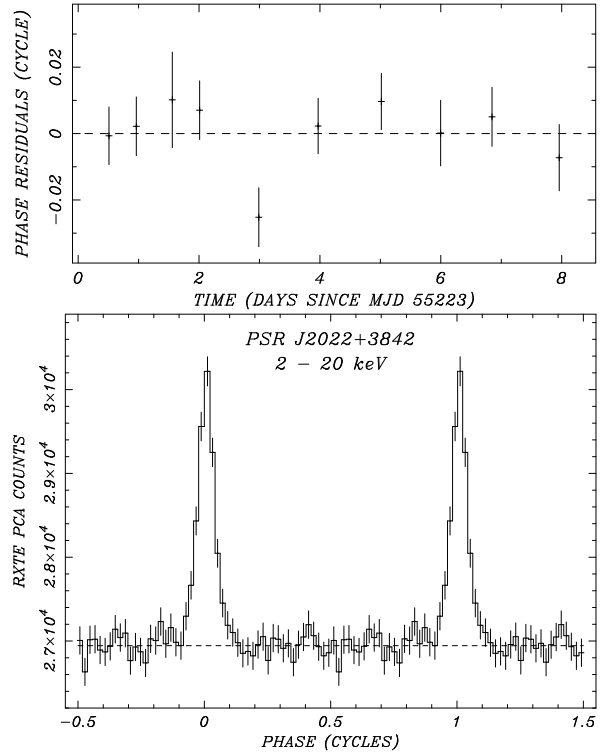

\hspace{0.1in}
\begin{center}
\includegraphics[height=0.9\linewidth,angle=270,clip=true]{psr2022_pca_toa_v3.ps}
\hfill
\includegraphics[height=0.9\linewidth,angle=270,clip=true]{psr2022_pca_fold_v3.ps}
\end{center}
\caption{\xte\ timing results for \psr\ in the 2--20~keV X-ray band.
{\em Top:\/} Phase residuals after fitting the phase-connected
quadratic ephemeris given in Table~\ref{tab:ephem} to the set of PCA
observations as described in \S\ref{sec:xtetim}. On average, individual TOA
measurements were obtained in 17 ks-PCU of exposure. {\em Bottom:\/} The 
pulse profile of the \xte\ data folded on the best-fit ephemeris. Two cycles
are shown for clarity.}
\label{fig:xtepulse}
\end{figure}

The \xte\ observations of \cxos\ distributed over the 8~days were
clustered into 10 groups of two, three, or four adjacent 96-min orbits,
with large gaps in between.  For each of these groups, we were able to
extract the period and phase of the signal with sufficiently small
uncertainties to maintain cycle counts between fitted epochs. 
Pulse profiles were generated using the optimum period derived from
the $Z^2_n$ test \citep{buc83} with $n=10$ harmonics to allow for
the sharp profile. The
resulting profiles were cross correlated, shifted, and summed to
generate a master pulse profile template.  Individual profiles were
then cross-correlated with the template to determine the arrival time 
(time of phase zero) and its uncertainty at each epoch.

The derived TOAs were iteratively fitted to a quadratic ephemeris
using the TEMPO software. We started by fitting TOAs from the first
three most closely spaced TOAs to a linear solution, and then
iteratively added the next TOA. At each step we found that the new TOA
would match to $<0.02$ cycles the predicted phase derived from the
previous set.  The resulting ephemeris referenced to the mid-point of
the observation is presented in Table~\ref{tab:ephem} and the phase
residuals are shown in Figure~\ref{fig:xtepulse}.  The residuals are
all less than 0.02 cycles and appear to be random within their
statistical uncertainties, with the exception of a single $3\sigma$
data point ($\Delta \phi =0.025\pm0.01$) which an investigation
finds no reason to exclude.

\begin{deluxetable}{ll}
\tabletypesize{\small}
\tablecaption{\label{tab:ephem}Properties of \psr }
\tablehead{
\colhead{Parameter}   &
\colhead{Value}   }
\startdata                                       
\multicolumn{2}{c}{\em X-ray Measurements} \\
R.A. (J2000)\tablenotemark{a}\dotfill          & $20^{\rm h}22^{\rm m}21\fs689(6)$\\
Decl. (J2000)\tablenotemark{a}\dotfill         & $+38\arcdeg42'14\farcs82(7)$    \\
Epoch (MJD TDB)\tablenotemark{b}\dotfill                        & 55227.00000027 \\  
Period, $P$ (ms)\tablenotemark{b}\dotfill                       & 24.2877561082(84)  \\
Period derivative, $\dot P$\tablenotemark{b}\dotfill            & $4.3064(93)\times10^{-14}$       \\
Range of timing solution (MJD)\tablenotemark{b}\dotfill         & 55223--55231       \\
\multicolumn{2}{c}{\em Radio Measurements} \\
Long-term average $\dot P$\dotfill	& $4.3192(27)\times10^{-14}$ \\
Range of $\dot P$ fit (MJD)\dotfill 	& 54957--55469	\\
Dispersion measure, DM (pc-cm$^{-3}$)\dotfill	& $429.1\pm0.5$	\\
Scatter-braodening timescale, $\tau^{\rm scatt}_{\rm 1\,GHz}$ (ms)\dotfill 	& 55(7) \\
Flux density, $S_{\rm 2\,GHz}$ ($\mu$Jy)\dotfill	& 60	\\
\multicolumn{2}{c}{\em Derived Parameters} \\
Characteristic age, $\tau_c$ (kyr)\dotfill      & 8.9                       \\
Spin-down luminosity, $\dot E$ (erg\,s$^{-1}$)\dotfill & $1.19\times10^{38}$   \\
Surface dipole magnetic field, $B_s$ (G)\dotfill & $1.03\times10^{12}$
\enddata
\tablecomments{\footnotesize $1\sigma$ uncertainties given in parentheses.}
\tablenotetext{a}{\footnotesize \chandra\ ACIS-S3 position registered using 
2MASS objects (see text).}
\tablenotetext{b}{\footnotesize \xte\ phase-connected ephemeris.}
\end{deluxetable}

Figure~\ref{fig:xtepulse} displays the pulse profile using all of the
2--20~keV data folded on the final ephemeris.  It has a single
symmetric peak that is triangular in shape and narrow, with a FWHM of
0.06 of a full cycle.  The measured pulsed emission corresponds to 0.91\%
of the total PCU countrate; however, as will be shown below, the intrinsic
signal is nearly 100\% pulsed compared to the flux derived from the
spectrum of \cxos.  We see no energy dependence of the pulse profile
when subdividing the 2--20~keV \xte\ energy range.

\subsection{\xte\ Spectral Analysis}
\label{sec:xtespec}

The pulsed-flux spectrum of \psr\ can be isolated through
phase-resolved spectroscopy. We used the FTOOLS {\tt fasebin}
software to construct phase-dependent spectra based on the ephemeris
of Table~\ref{tab:ephem}.  We combined spectra from all observation
intervals using only data recorded in the top Xenon layer of PCU\#2;
this PCU was (uniquely) used for all observation intervals over the
8~days. Similarly, we averaged the standard PCA responses generated
for each epoch.  In fitting the pulsed flux, the unpulsed emission
provides a near perfect background estimate.  The spectra were
accumulated in two phase intervals representing the off-peak
($0.1<\phi_{\rm off}\leq0.88$) and on-peak ($0.88<\phi_{\rm on}\leq1.1$) 
emission and fitted using XSPEC. The fits were performed in the
2--16~keV range, above which the background dominates.

\begin{figure}[t]
\centerline{
\includegraphics[height=0.95\linewidth,angle=270,clip=true]{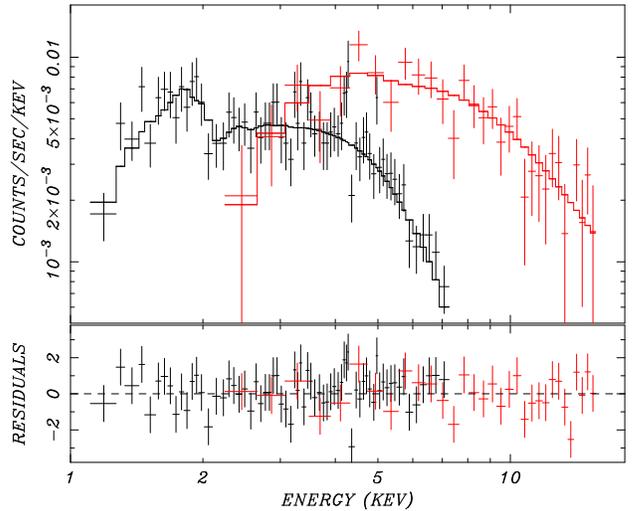}
}
\caption{The \chandra\ ACIS spectrum of \cxos\ ({\it black}) and the \xte\ PCA spectrum 
of the pulsed emission from \psr\ ({\it red}), fitted jointly to an
absorbed power-law model, with independent normalizations. The solid
line shows the best-fit model (see text \S\ref{sec:xtespec}). The pulsed \xte\ flux %from \psr\
is obtained by subtracting the off-peak spectrum from the on-peak
spectrum. Residuals from the best fit are shown in units of standard
deviation.}
\label{fig:chandraspectra}
\end{figure}

Based on the \chandra\ spectrum, we fitted an absorbed power-law model
with the interstellar absorption held fixed at $N_{\rm H} =
1.6\times10^{22}$~cm$^{-2}$; leaving $N_{\rm H}$ unconstrained
results in a larger uncertainty in $\Gamma$.  The resulting best-fit
photon index is $\Gamma = 1.1\pm0.2$ with $\chi^2 = 27.9$ for 30 DoF.
The absorbed 2--10~keV flux for the pulsed emission is $5.4 \times
10^{-13}$~erg~cm$^{-2}$~s$^{-1}$, which represents all of the
point-source flux measured from \cxos, indicating that its
intrinsic pulsed fraction is essentially 100\%. In
Figure~\ref{fig:chandraspectra}, we show the result of a joint fit to
the \xte\ pulsed emission from \psr\ and
\cxos\ in the 0.5--16~keV band, leaving only their normalizations free.  
The best fit parameters are $N_{\rm H} = (1.7\pm0.3)\times
10^{22}$~cm$^{-2}$ and $\Gamma = 1.0\pm0.2$ with $\chi^2 = 99.3$ for
100 DoF.  There is no indication of spectral curvature in the fitted
energy range.
The independently measured absorbed 2--10~keV flux is the same for both 
\cxos\ and the pulsed emission from \psr, namely $F_{X} = 5.3 \times
10^{-13}$~erg~cm$^{-2}$~s$^{-1}$. We adopt the results of the joint
spectral fit as our final reported values.

\section{Discussion}
\label{sec:discuss}
There can be little doubt that \cxos\ and \psr\ are one and the same:
the non-thermal X-ray source lies at the center of \snr, lacks an
optical counterpart, and anchors nebular emission consistent with a
PWN. Most importantly, the pulsed flux detected using \xte\ accounts
for all of the \chandra\ emission and no more. These properties are
consistent with pulsars producing broadband magnetospheric emission
powered by rotational energy losses. \chandra\ astrometry thus locates
the pulsar to sub-arcsec precision.

\psr\ is outstanding in several respects. It is the second-most energetic 
Galactic pulsar known, after the Crab pulsar, and the fourth overall,
taking into account two LMC pulsars, in N157B (PSR~J0537$-$6910,
with $\dot E = 4.9\times 10^{38}$~erg~s$^{-1}$ and $P=16$~ms)
and in SNR 0540$-$69 (PSR~J0540$-$6919, with $\dot E = 1.5\times
10^{38}$~erg~s$^{-1}$ and $P=50$~ms).  However, it is among the
least efficient at
converting its spin-down luminosity into X-rays\footnote{For 
ease of comparison with the compilation of \citet{2008AIPC..983..171K}, the 
flux, luminosity, and efficiency in the 0.5--8 keV band are 
$F_{\rm PSR} = 4.0\times10^{-13}$~ergs~cm$^{-2}$~s$^{-1}$,
$L_{\rm X} = 6.5\times10^{33}\,D^2_{10}$~ergs~s$^{-1}$, and
$L_{\rm X}/\dot E = 5.5\times 10^{-5}\,D^2_{10}$.
}, with $\eta \equiv
L_{\rm 2-10\,keV} /\dot E = 5.9\times10^{-5}\, D^2_{10}$.
%, similar to the much less energetic Vela pulsar. 
%($\dot E = 6.9\times10^{36}$~erg~s$^{-1}$, $P=89$~ms, PSR~J0835$-$4510). 
The flat pulsed
spectrum also sets it apart from the typical young, energetic pulsar,
deviating significantly from the observed trend $\Gamma \propto
1/\sqrt{\dot E}$
\citep[][2--10~keV]{got03}, which predicts an index of $\Gamma=1.8$. 
Flat power laws have primarily been found among pulsars with
\hess\ counterparts \citep[e.g., PSR~J1838$-$0655;][]{gh08}. 

\begin{figure}[t]
\hspace{0.1in}
\includegraphics[height=0.9\linewidth,angle=270,clip=true]{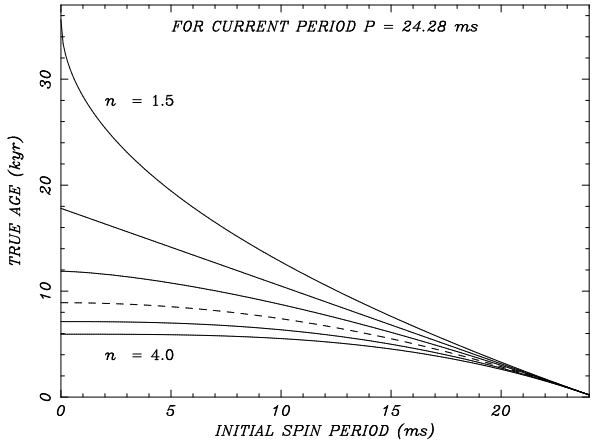}
\caption{The true age $\tau$ of \psr\ as a function of
its spin period at birth, $P_0$, according to Equation~(\ref{eq:psrage});
$n$ is the braking index.  The curves correspond,
from top to bottom, to $n=1.5$, 2.0, 2.5, 3.0 (dashed), 3.5, and 4.0, converging at
the present-day period, $P = 24.28$~ms. 
The associated upper-limits on the pulsar age (for $P_0$ much smaller
than $P$) are $\sim 36$, 18, 12, 9, 7, and 6 kyr, respectively. 
}
\label{fig:psrage}
\end{figure}

\psr\ is also the second-most rapidly rotating young pulsar after
the LMC pulsar J0537$-$6910, and the shortest-period radio-bright
pulsar known. Its radio properties are thus potentially interesting
probes of the elusive radio emission mechanism in pulsars, but
detailed studies will be hampered by this object's faintness and the
interstellar propagation effects imposed by its great distance. The
short rotation period also engenders uncertainty in the pulsar's
(and thus the supernova remnant's) age: the characteristic age
$\tau_c = P/2\dot P = 8.9$~kyr approximates the true age $\tau$ only when $P_0
\ll P$, where $P_0$ is the spin period at birth,

\begin{equation}
\label{eq:psrage}
\tau = \frac{P}{(n-1)\dot{P}} \left[1-\left(\frac{P_0}{P}\right)^{n-1}\right],
\end{equation}

\noindent where $n$ is the so-called ``braking index.''
For a birth period of 16 ms, the shortest known among young pulsars, the
true age decreases to $\approx 5$~kyr, assuming spin-down due to
magnetic dipole braking ($n=3$). A
countering effect, however, is spindown that deviates from the
standard dipole assumption: the measured braking indices for several
young pulsars are found to lie within the range $2.0 \lesssim n <
3.0$ (see, e.g., \citealp{1997MNRAS.288.1049M}). For example, as
shown in Figure~\ref{fig:psrage}, $P_0 = 10$~ms for $n = 2$ would
imply an age of 11 kyr. For any assumed value of $n$, an upper limit
to the age of the pulsar, and thus the remnant, is provided in the
limit $P_0 \rightarrow 0$. A braking index well below $n = 1.5$ would be
highly unusual, so that $\sim 40$~kyr is a reasonable upper limit on
the age of \snr.

\psr\ lacks the bright X-ray PWN expected from an energetic pulsar:
it is only the second example (of some 20) of a pulsar with $\dot E
\simgt 4\times 10^{36}$~erg~s$^{-1}$ unaccompanied by a PWN of
comparable brightness, $F_{\rm PWN}/F_{\rm PSR}\simgt 1$, defying
the trend presented in Gotthelf (2004). As such, it is similar  to
\psrb\ (Torii \etal\ 1998), the 69~ms pulsar with $\dot E = 1.6
\times 10^{37}$~erg~s$^{-1}$, $B_s = 3.1 \times 10^{12}$~G, and
$\tau_c = 8.1$~kyr.  %Give the current state of understanding of PWN, 
The underluminous PWNe in these cases remain unexplained. In
other respects, however, the X-ray PWN around \psr\ may not be so
unusual. The semi-major axis of the elliptical elongation around the
unresolved source is approximately $6\arcsec$ in length,
representing a physical dimension of $9\times10^{17} \ D_{10}$~cm.
For comparison, the Vela PWN's X-ray ``outer arc''
lies at a distance $1\times10^{17}$~cm from the pulsar in the geometric model of
\citet{2001ApJ...556..380H}.
%This is entirely consistent with the size of the ``inner arc''
%associated with the Vela pulsar, $5.2\times10^{16}$~cm, when scaled
%by the relative spin-down luminosities of the two pulsars, $\dot
%E_{\rm J2022}/\dot E_{\rm Vela} = 17.2$. If this is not simply a
%numerical coincidence, the implication is that the arc-like features
%occur at a pressure-balance point between each pulsar's Poynting
%flux and the ambient medium (the pressure of which must then also be
%comparable for the two pulsars). Deeper high-resolution imaging of
%the PWN around \psr\ will determine whether these tentative
%conclusions have merit. 

If the extended emission around \psr\ does arise from axisymmetric
features, the \chandra\ image constrains---in analogy with the Crab,
Vela, and others---the orientation of the pulsar's spin axis: in
inclination, to $\simeq 50^\circ$, and on the sky, to the symmetry
axis at position angle $-64^\circ$ North through East, orthogonal to
the long direction of the diffuse X-ray emission and the ridge
linking the radio lobes. The apparent alignment of spin axes and
proper motion vectors in some of these same pulsars offers a
testable prediction for the motion of \psr. The spin inclination
relative to the line of sight, meanwhile, may help elucidate the
role of viewing geometry (e.g., relative to a narrow
magnetospheric emission beam) in {\em i)\/} producing simultaneously a narrow
X-ray pulse but an unexpectedly broad radio pulse, a reversal of
their typical relationship, and {\em ii)\/} determining whether the observed
neutron star spectrum is primarily thermal or nonthermal: the flat
spectrum of \psr\ departs markedly from the dominant thermal
emissions of the central stars in the Vela and DA~495 SNRs.

We have already alluded to some of the notable similarities between
\snr\ and DA~495, both of which can now unambiguously be
characterized as radio PWNe. Both are very bright radio sources
without a well-defined outer boundary. Both manifest a bipolar
structure and have a steep radio spectrum, unusual for a PWN. DA 495
has a break in its radio spectrum at roughly 1 GHz, and \snr\ may
have one below that as well. Both show very faint X-ray emission
compared to their radio emission, and a small X-ray nebula compared
to the radio nebula. The conclusions arrived at by 
ASL+08 for DA~495 may thus also
apply to \snr. For example, based on the radio/X-ray PWN size ratio
of DA~495, $\gtrsim 25$, and other lines of argument \citep[see
also][]{2008ApJ...687..516K}, ASL+08 suggested that the pulsar wind
energizing DA~495 has a high magnetization factor---i.e., it carries
electromagnetic flux and particle flux at roughly comparable
levels---in contrast with the Crab, which has a strongly
particle-dominated wind, but similar to independent assessments of
the wind properties of the Vela pulsar. Moreover, ASL+08 speculated
that, because particle-dominated winds are necessary for efficient
conversion of wind luminosity to synchrotron luminosity, PWNe in
which Poynting flux is an important wind component may be those with
dim X-ray PWNe. 

The PWN size ratio for \snr\ is $\simeq 20$, comparable to DA~495
and nearly two orders of magnitude greater than that for the Crab
Nebula. (At a distance of 10 kpc, G76.9+1.0 may be the largest PWN
known in our Galaxy, with a physical size of $29 \times 35$~pc,
bigger than or comparable to MSH 15$-$57, which is at least 25 pc in
diameter, \citealp{2000ApJ...542..380G}, and so far believed to be
the largest Galactic PWN.) If our prior conclusions hold, the wind
from \psr\ should be highly magnetized well beyond its termination
shock, contributing to the low X-ray conversion efficiency of its
PWN. Why this might be so is an open question.

The overarching conclusion of ASL+08 and \citet{2008ApJ...687..516K}
was that DA~495 is likely to be a PWN of advanced age ($\sim 20$~kyr), but having
evolved without significant interaction with the ambient medium. All
of the characteristics we find for \snr\ similarly indicate a rather
old object, yet \psr\ has all the characteristics of a young object: a spin-down age
of 20~kyr would require both $n < 2$ and very high spin at birth, $P_0 \lesssim 5$~ms.
We do not have a ready explanation for this apparent discrepancy,
which highlights the importance of having uncovered the spin and
energetic properties of the central pulsar. 

\snr\ and its pulsar, J2022+3842, are clearly unusual and require further
investigation. Additional multi-wavelength observations (e.g., an X-ray 
search for the SNR shell, to aid in constraining the system's age) 
may be fruitful in providing important components for their understanding.  
Based on its spin-down luminosity, and given the sub-arcsec localization 
and available timing information, \psr\ is a good candidate for a search
for gamma-ray pulsations using {\em Fermi\/} data. The absence of a bright
PWN, however, makes it an unlikely TeV target, even though the
spectrum of the pulsed X-ray emission suggests otherwise.

\acknowledgements

Support for this work was provided by the National Aeronautics and
Space Administration through \chandra\ Award Number GO5-6077Z issued
by the \chandra\ X-ray Observatory Center, which is operated by the
Smithsonian Astrophysical Observatory for and on behalf of 
NASA under contract NAS8-03060.
The National Radio Astronomy Observatory is a facility of the
National Science Foundation operated under cooperative agreement by
Associated Universities, Inc.
We have also made use of \xte\ data provided by the High Energy
Astrophysics Archive at NASA's Goddard Space Flight Center, as well as
data products from the Two Micron All Sky Survey, a joint project of the 
University of Massachusetts and the Infrared Processing and Analysis Center/Caltech, 
funded by NASA and the National Science Foundation.
SSH acknowledges support by the Natural Sciences and Engineering
Research Council of Canada (NSERC) and the Canada Research Chairs
program.

\end{document}